\documentclass{INTERSPEECH2023}

\interspeechcameraready

\usepackage{color,soul}
\usepackage{caption}
\usepackage{subcaption}
\usepackage{multirow}
\usepackage{microtype}
\usepackage{hyperref}
\definecolor{abhi-color}{rgb}{0, 0, 0}
\newcommand{\abhi}[1]{\textcolor{abhi-color}{#1}}
\newcommand{\model}{\textsc{adaptermix}}
\title{\model{}: Exploring the Efficacy of \textit{Mixture of Adapters} for Low-Resource TTS Adaptation}

\let\realcite\cite
\renewcommand{\cite}[1]{\ifx&#1&\hl{[?]}\else\realcite{#1}\fi}

\author{Soujanya Poria}
\date{February 2023}
\name{
Ambuj Mehrish$^1$, Abhinav Ramesh Kashyap$^2$, Li Yingting$^3$, Navonil Majumder$^1$, Soujanya Poria$^1$}
\address{
  $^1$Singapore University of Technology and Design, Singapore\\
  $^2$ASUS Intelligent Cloud Services (AICS) Singapore, \\
  $^3$Beijing University of Posts and Telecommunications, China}
\email{ambuj\_mehrish@sutd.edu.sg, abhinav\_kashyap@asus.com, cindyyting@bupt.edu.cn, navonil\_majumder@sutd.edu.sg, sporia@sutd.edu.sg}

\begin{document}

\maketitle
 
\begin{abstract}
 There are significant challenges for speaker adaptation in text-to-speech for languages that are not widely spoken or for speakers with accents or dialects that are not well-represented in the training data. To address this issue, we propose the use of the "mixture of adapters" method. This approach involves adding multiple adapters within a backbone-model layer to learn the unique characteristics of different speakers.
Our approach outperforms the baseline, with a noticeable improvement of 5\% observed in speaker preference tests when using only one minute of data for each new speaker. Moreover, following the adapter paradigm, we fine-tune only the adapter parameters (11\% of the total model parameters). This is a significant achievement in parameter-efficient speaker adaptation, and one of the first models of its kind.
Overall, our proposed approach offers a promising solution to the speech synthesis techniques, particularly for adapting to speakers from diverse backgrounds.

\end{abstract}

\section{Introduction}
\label{sec:intro}


One of the key aspects of text-to-speech  (TTS) technology is the ability to capture the unique acoustic mannerisms of a given speaker, which can include characteristics such as accent, intonation, rhythm, and other vocal traits that are associated with a speaker's identity~\cite{mehrish2023review}. This can be especially important in applications where the speaker's identity is a key factor, such as in voice assistants or interactive voice response systems. Thus, capturing these vocal idiosyncrasies in the generated speech is challenging and often requires many hours of reference speech samples from the speaker. Performing TTS tasks using large reference samples could be infeasible due to various reasons, such as limited memory budget, privacy concerns, or logistical issues. To address this challenge, we suggest a low-resource TTS approach that utilizes reference samples no longer than 1 minute. This strategy will help overcome the aforementioned limitations and enable efficient TTS performance with minimal resources.


TTS is being integrated into a wide range of applications, from virtual assistants to audiobooks, making it more accessible and useful than ever before. As text-to-speech technology continues to evolve, it has the potential to revolutionize communication and accessibility for people with disabilities or language barriers. One of the biggest challenges in TTS research is improving the naturalness and expressiveness of speech \cite{alimproving,kim21n_interspeech,b2022diffusion}. This involves using algorithms to convert written text into spoken words, and there are many techniques available to make the resulting speech sound more natural and expressive. For example, prosody modelling \cite{raitio2022hierarchical,chien2021hierarchical} can help to convey intonation, stress, and rhythm, while neural network-based models such as \cite{shen2018natural,renfastspeech} can produce speech with more natural sounding timbre and articulation.


Few-shot \emph{speaker adaptation} is a challenging task in TTS, which aims to personalize synthesized speech to match the characteristics of a specific speaker. This can involve modifying the acoustic model of the TTS to adjust for factors such as pitch, tone, and pronunciation \cite{cooper2020zero,casanova2022yourtts,wu2022adaspeech}. Adaptive TTS models can be used to achieve few-shot speaker adaptation through methods such as pre-trained speaker embedding models with a small amount of reference speech \cite{casanova2022yourtts} or fine-tuning multi-speaker TTS. However, fine-tuning requires a large amount of training data to avoid over-fitting and catastrophic forgetting \cite{hemati2021continual}.

In this paper, we propose a low-resource speaker adaptation approach based on the mixture of adapters (MoA) \cite{wang2022adamix,jiang2023mixphm,chronopoulou2023adaptersoup}. MoA has been widely used in NLP tasks such as text  \cite{wang2022adamix}, question answering  \cite{jiang2023mixphm}, and machine translation \cite{baziotis2022multilingual}, by fine-tuning pre-trained models with minimal adapter parameters. In TTS, MoA can train a parameter-efficient module on a small amount of data, enabling fine-tuning of a pre-trained TTS model for a new speaker without forgetting previous knowledge. MoA's multiple adapter modules capture fine-grained information about the speaker's speech  such as prosody, speaking rate or accent and adapt the acoustic model more effectively than traditional fine-tuning, saving time and computational resources while maintaining or improving performance.

MoA has several advantages over traditional fine-tuning approaches, such as faster training time, better generalization to new tasks, and improved robustness to domain shifts. However, MoA also has some limitations, such as the need for many adapters to achieve good performance, the risk of overfitting on small datasets, and the difficulty of interpreting the adapter weights. While MoA is a promising technique for improving the performance of TTS systems with minimal resources, its use is likely to increase in the future.

To this end, we take a two-phase training approach: i) train a transformer-based encoder-decoder TTS model on a large TTS dataset, LibriTTS \cite{zen2019libritts} with $100$h of clean speech samples from $251$ speakers; ii) adapt this backbone model to work with short duration speech samples. The first phase is meant to learn the TTS task, traditionally by optimizing all the parameters in the network. Whereas, the second phase adds only a small fraction of the parameters, in the form of Adapters~\cite{houlsby2019parameter}, only to optimize them to work with shorter reference speech samples. As compared to the recently proposed by Hsieh~\cite{morioka2022residual}, which trains an adapter for each speaker, our approach is much more scalable to a large number of speakers as the adapters are shared among the speakers. To learn the key lower-dimensional vocal qualities, from the speech samples in the context of the target text, we add multiple parallel adapters to the transformer decoder and aggregate their outputs, akin to \emph{mixture of experts}~\cite{zhou2022mixture}. We call our setup \model{}.

\section{Related Work}
\label{sec:related}
There are various approaches to few-shot speaker adaptation for TTS systems \cite{chen2018sample, kons2019high, arik2018neural}. The main idea behind these methods is to first train a TTS model \cite{ren2020fastspeech, shen2018natural} on a large dataset of speech from multiple speakers \cite{zen2019libritts, Yamagishi2019CSTRVC}. This enables the model to learn general speech and language patterns. Subsequently, the model is fine-tuned on a smaller dataset of speech from a new speaker \cite{casanova2022yourtts}. One significant advantage of using a pretrained model for few-shot adaptation is that it can quickly adapt to a new speaker with minimal data by leveraging the knowledge learned from the large dataset. This is particularly useful when collecting significant amounts of data for each new speaker is impractical or infeasible \cite{cooper2020zero}. Numerous techniques have been proposed for few-shot speaker adaptation using pretrained TTS models \cite{casanova2022yourtts,wu2022adaspeech}, including meta-learning \cite{huang2022meta}, transfer learning \cite{jia2018transfer, neekhara2021adapting}, and domain adaptation \cite{zhang2022tdass}. These approaches aim to enhance adaptation performance and reduce the data requirements for successful adaptation. A notable example of such an approach is Adaspeech \cite{chen2021adaspeech}, which achieves efficient adaptation and edge-affordable memory storage by incorporating conditional layer normalization into the decoder.
Adapters-based approaches have shown remarkable performance in low-resource TTS by efficiently using only a fraction of parameters compared to fully fine-tuned models. In \cite{morioka2022residual}, the residual adapter was recently introduced in TTS, but it trains a specific adapter for each target speaker, limiting its scalability as the number of speakers increases. In our approach, we also use residual adapters in the decoder but with $N$ adapters in each layer, which are shared by all target speakers, and a routing algorithm selects tokens during training. This design significantly reduces the number of parameters and allows scalability for larger datasets compared to \cite{morioka2022residual}.

\section{Proposed Method}
\label{sec:proposed}
We first pretrain a multi-speaker Transformer TTS model \cite{li2019neural} on LibraTTS corpus~\cite{zen2019libritts} and then adapt the model to new speakers by only training the \emph{mixture of adapters} module inserted into the decoder layer of the pretrain model. In this section, we will briefly discuss the architecture of the Transformer TTS backbone model, followed by \model{}.
\vspace{-0.5em}
\subsection{Transformer TTS backbone architecture}

Transformer~\cite{vaswani2017attention} architecture has been successfully applied to TTS systems~\cite{li2019neural,ren2020fastspeech}, consisting of a text encoder, duration and pitch predictors, decoder, and vocoder. The text encoder uses self-attention and feedforward neural networks to generate a new representation of the input text. The duration predictor and pitch predictor estimate phoneme lengths and fundamental frequency, respectively. The decoder takes the encoded text, predicted phoneme durations, and predicted F0 values to generate spectrograms, which are converted into a time-domain waveform by a vocoder. Additional Postnet and linear projection layers are used to enhance speech quality.

\subsection{AdapterMix}
\begin{figure*}[t]
  \begin{subfigure}[b]{0.5\textwidth}
    \centering
    \includegraphics[height=0.35\textheight, width=0.45\columnwidth]{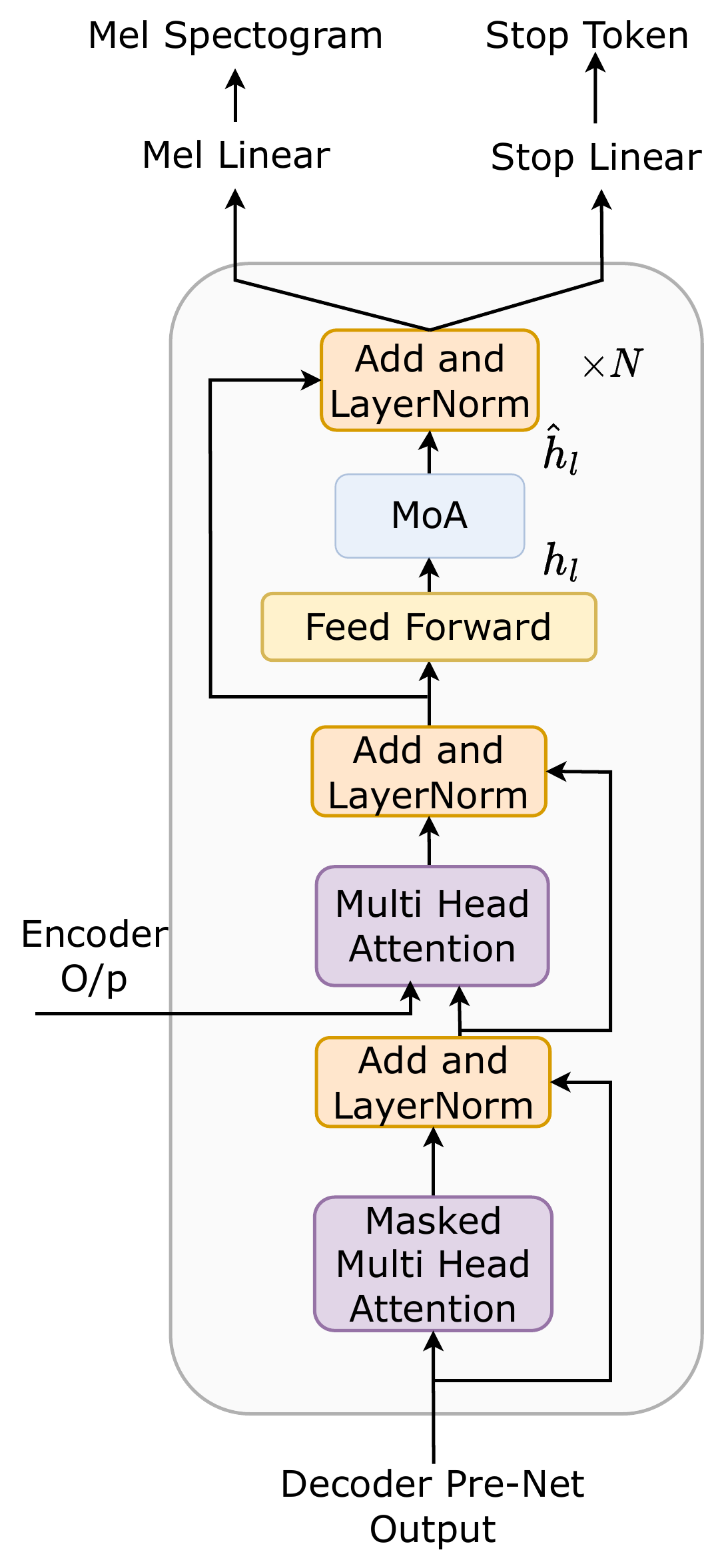}
       \caption{}
    \label{fig:overall_of_adaptermix}
  \end{subfigure}%
  \begin{subfigure}[b]{0.52\textwidth}
    \centering
    \includegraphics[width=\columnwidth]{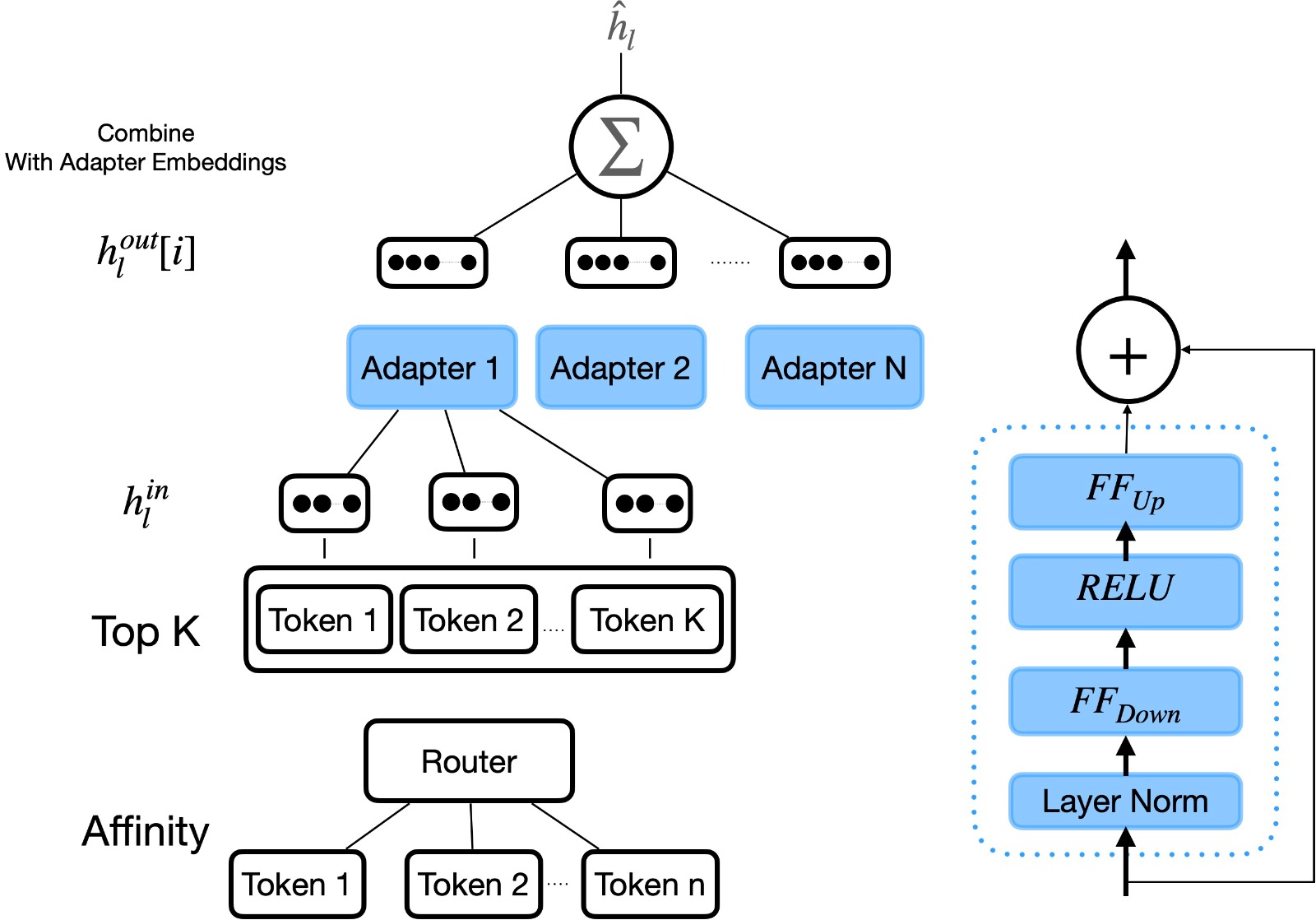}
   \caption{}
    \label{fig:MoA}
  \end{subfigure}
  \caption{(a) Transformer TTS \cite{li2019neural} decoder architecture with one MoA module. (b) The MoA module comprises $N$ residual adapters (left) \abhi{Every adapter chooses $k$ closest tokens and processes it. The same token can be processed by multiple adapters. The outputs of the adapters are combined}. Additionally, the architecture of the standard residual adapter is illustrated on the right in the same diagram. }
  \label{fig:mainfig}
\end{figure*}

We adapt the pretrained transformer TTS backbone to a new target speaker using mixture of adapter modules inserted into every layer of the decoder after the feed-forward sub-layer, as shown in Figure \ref{fig:overall_of_adaptermix}. \emph{Mixture of adapters} module consists of $N$ lightweight neural modules and a routing algorithm for independently selecting top-$k$ tokens for individual adapters. By selecting tokens that are highly relevant to speaker identity, we can effectively adapt the model to new speakers while minimizing the need for extensive fine-tuning. Whereas the individual adapters are meant to capture complementary characteristics of the tokens, in context of the speaker, that are relevant to TTS.

As illustrated in Figure \ref{fig:MoA}, \model{} consists of $N$ residual adapters and a routing algorithm. The architecture of the residual adapter is also shown in Figure \ref{fig:MoA}. Each residual adapter first applies layer normalization \cite{ba2016layer} to a $d_{model}$ dimensional input vector $h_{l} \in \mathbf{R}^{d_{model}}$ (subscript $l$ represents $l^{th}$ decoder layer). Following normalization, the output is projected down to a bottleneck dimension $r$, followed by a nonlinearity (ReLU \cite{krizhevsky2017imagenet}), up projection to the model dimension $d_{model}$ and finally a residual connection. The residual adapter can be described in the following way:
\begin{equation}
    \hat{h}_{l} = h_{l} + ReLU(LayerNorm(h_{l}) W_{down}) W_{up}
    \label{eq:RA}
\end{equation}
where $W_{down} \in \mathbf{R}^{d_{model} \times r}$ and $W_{up} \in \mathbf{R}^{r\times d_{model}}$ are the down and up projection weights. 


\model{} use an expert choice routing strategy similar to \cite{zhou2022mixture}, where top-$k$ tokens are selected independently for each adapter. Depending on the number of adapters $N$, capacity $c$ (hyperparameter), and length of the sequence $n$, $k$ is determined dynamically as $k = \frac{n \times c}{N}$. 
As a result, tokens can be allocated to a variable number of adapters, enabling flexible allocation. The token-to-adapter affinity score $S \in \mathbf{R}^{n \times N}$ is computed using standard softmax operation between input $h_{l}$ and $W_{g} \in \mathbf{R}^{d_{model} \times N}$, where $W_{g}$ denotes adapters embeddings. Following equations can be used to explain routing

\begin{equation}
    S = Softmax(h_{l},W_{g})
\end{equation}
Top $k$ largest entries of each row in $S^{T}$ are selected as an input to the $N$ adapters:
\begin{equation}
\label{eq:perm}
    G,I = TopK(S^{T},k).
\end{equation}
The input to each adapter $h_{l}^{in}$ is obtained by permutation matrix $ P = Onehot(I) $ is the one hot version of $I$. This helps us to pick up the appropriate token representation for a given adapter as:
\begin{equation}
    h_{l}^{in} = P h_{l}.
\end{equation}

The output of $i^{th}$ the adapter $h_{l}^{out}[i]$ is computed using Equation \ref{eq:RA}, where $h_{l}^{in}[i]$ is input to each adapter. \abhi{Finally, all the adapter outputs are combined with $G$ and  $P$ from Equation \ref{eq:perm} using Einstein summation (einsum) operations}:
\begin{equation}
    \hat{h}_{l}= h_l + \sum_{i} P G h_{l}^{out}[i].
\end{equation}
\section{Experiments}
\label{sec:exp}
\subsection{Baselines}
As a common baseline, we evaluated \model{} against two other methods: full fine-tuning and residual adapter \cite{morioka2022residual}. The residual adapter-based baseline is hereafter referred to as \texttt{Adapter}. During full fine-tuning, all parameters of the backbone model are updated, while for \texttt{Adapter}, only adapter parameters are updated, keeping the backbone frozen.

\subsection{Training setup}
The multi-speaker backbone model utilized a Transformer architecture, specifically a Transformer-TTS model~\footnote{The source codes of AdapterMix along with the checkpoints are publicly available at \url{https://github.com/declare-lab/adapter-mix}.}, which consisted of 4 encoders and 6 decoder layers, with a hidden state dimension of $d_{model}=256$. This model also incorporated other speech synthesis modules, including post-net, pre-net, and variance adapter, as described in Section \ref{sec:proposed}.

To pretrain the backbone model, we used the train-clean-100 split of the multi-speaker English LibriTTS corpus \cite{zen2019libritts}, which contains 100 hours of $24$ kHz English speech from 251 speakers. We downsampled the speech to $22.05$ kHz and trained the model for $900$k steps using the Adam optimizer. Overall, the Transformer-TTS model had $3.6$M parameters 

We randomly selected ten speakers (five male and five female) from the CSTR VCTK corpus \cite{Yamagishi2019CSTRVC} for our speaker adaptation experiments. We then divided the selected speakers' utterances into three groups based on training duration: 1 min, 10 min, and 15 min. We downsampled the speech to 22.05 kHz, used the Adam optimizer for training, and trained all models on a single NVIDIA Tesla A6000 GPU. To ensure a fair comparison between the baselines (Finetune, \texttt{Adapter}) and MoA (\model{}), we trained each one of them for 10k steps. We applied a warm-up period of 4000 steps and performed learning rate annealing at 6000, 7000, and 8000 steps, with an annealing rate of 0.3. All models were trained with a batch size of 64, except in the case of the 1-min training duration, where we used a batch size of 16.
\begin{table*}[h]
    \centering
    \caption{Subjective (MOS) and objective comparison among full Finetune, \texttt{Adapter}, and \model{}. }
    \resizebox{\textwidth}{!}{%
    \begin{tabular}{@{}llccccccccccccc@{}} 
    \toprule
    \multirow{2}{*}{Method (\#param \%)}& \multicolumn{4}{c}{1min}& \multicolumn{4}{c}{10min} &\multicolumn{4}{c}{15min}\\
    \cmidrule(lr){2-5}\cmidrule(lr){6-9} \cmidrule(lr){10-13}
    {}& MOS$\uparrow$ & MCD$\downarrow$& WER$\downarrow$& Cosine Sim$\uparrow$ & MOS$\uparrow$ & MCD$\downarrow$& WER$\downarrow$& Cosine Sim$\uparrow$ & MOS$\uparrow$ & MCD$\downarrow$& WER$\downarrow$& Cosine Sim$\uparrow$\\
    \midrule
    {Ground Truth} & 4.01$\pm$ 0.33 & -  &0.19036  & - & 4.10 $\pm$ 0.41 & - &0.1853  & - & 4.17 $\pm$ 0.41  & - &0.1861 &-\\
    {Finetune} & 3.45$\pm$ 0.48 & 5.7450 & \bf 0.2489  & \bf 0.747 $\pm$ 0.0065  & 3.18 $\pm$ 0.29 &5.6453  &\bf 0.2228  & \bf 0.7362 $\pm$ 0.0069 & 3.54 $\pm$ 0.49  &5.7058 &\bf 0.2056 &\bf 0.7374 $\pm$ 0.0080\\
    {\texttt{Adapter} (1.57\%)} & 2.82 $\pm$ 0.52 &  \bf 5.2482 & 0.3911  & 0.6733$\pm$ 0.0072 & 3.13 $\pm$ 0.42  & \bf 5.2763  &0.2445  &0.7091 $\pm$ 0.0053 & 3.19 $\pm$ 0.42  &\bf 4.9430 &0.2731 &0.6957 $\pm$ 0.0062\\
    {\model{} (11.62\%)} & 3.33 $\pm$ 0.31 & 5.5943 & 0.2987  & 0.7037 $\pm$ 0.0119 & 3.66 $\pm$ 0.31 &5.4216  & 0.2270  &0.7324 $\pm$ 0.0060 & 3.53 $\pm$ 0.37  & 5.3641&0.2260 &0.7170 $\pm$ 0.0081\\\bottomrule
\end{tabular}}
    \label{tab:performance}
\end{table*}

During the adaptation process to a new speaker, we insert adapter modules (\model{} and \texttt{Adapter}) into the decoder layer of the backbone model, as shown in Figure \ref{fig:overall_of_adaptermix}. Unless stated otherwise, the bottleneck dimension $r$ is set to $128$. Target speakers may have different variance information, such as duration, pitch, energy, etc., than the speakers on which the backbone model is pretrained. To capture the variance of target speakers, we add a single residual adapter to the output of the variance adapter with $r=64$. We optimize only the adapter modules (\model{} and \texttt{Adapter}) for the target speaker, while keeping the entire backbone frozen, including batch normalization. We also fine-tune the target speaker's speaker embedding to ensure that the variance adapters are properly conditioned by the learned target speaker embedding, in addition to the adapter modules (\model{} and \texttt{Adapter}). By doing so, we can better capture the target speaker's unique characteristics and improve the overall quality of the synthesized speech.

\subsection{Objective Evaluation}
In this study, we evaluated the mel-spectrogram reconstruction capability of the synthesized speech using the mel-cepstrum distortion (MCD) \cite{toda2007voice}. To assess the intelligibility of the synthesized speech, we used the word error rate (WER). Additionally, we calculated an objective speaker similarity measure by computing the mean cosine similarity between the embeddings of the ground truth and synthetic speech samples, using a neural speaker embedding system called deep speaker \cite{li2017deep}. The embeddings used for computing cosine similarity had a dimension of $512$. Each speaker similarity test had a total of approximately $200$ utterances.
Although all the models produced similar MCD scores, Table \ref{tab:performance} highlights performance gaps between \model{} and the other two baselines. Notably, \model{} outperforms \texttt{Adapter} in terms of WER, but performs comparably in terms of MCD for all durations.
Thus, although \model{} scores slightly lower in mel-spectrogram reconstruction capability, it generates more intelligible speech than \texttt{Adapter}.

The cosine similarity scores of \model{} are comparable to those of the finetuned model and better than those of \texttt{Adapter} for all time intervals. For a model trained on $10$ minutes of data per speaker, \model{} achieves a cosine similarity score of $0.7324$, compared to $0.7362$ and $0.7091$ for the finetuned model and \texttt{Adapter}, respectively. This result indicates that \model{} has better speaker adaptation capabilities than the other baselines. In the objective evaluation, Fine-tuning has an edge over \model{}, which can be credited to the number of trainable parameters in the fine-tuned model. After conducting a subjective evaluation (refer to Section \ref{sec:subjeval}), it was found that \model{} performs as well as the fine-tuned model and, in some experiments, even outperforms it. These results demonstrate the effectiveness of \model{}, which utilizes only 11\% of the trainable parameters present in the full fine-tuned model.

\begin{figure}[h]
    \centering
    \includegraphics[width=0.9\columnwidth]{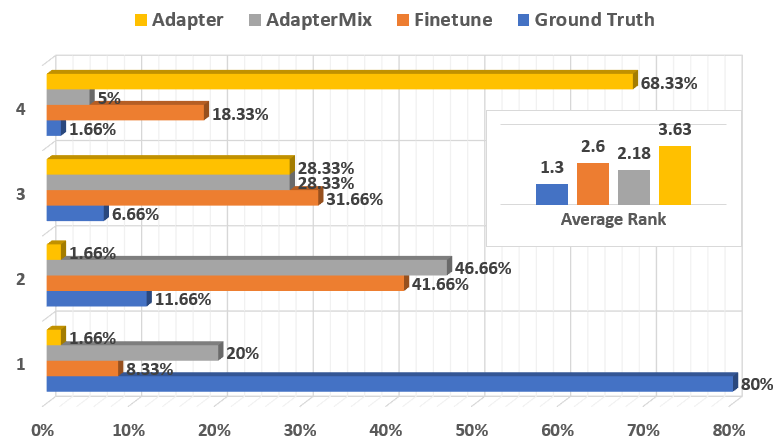}
    \caption{Ranking of the \% of speech samples synthesized by each model based on the MOS score.}
    \label{fig:rank}
\end{figure}

\subsection{Subjective Evaluation}
\label{sec:subjeval}

\begin{figure}[hb]
\centering
  \includegraphics[width=0.8\columnwidth]{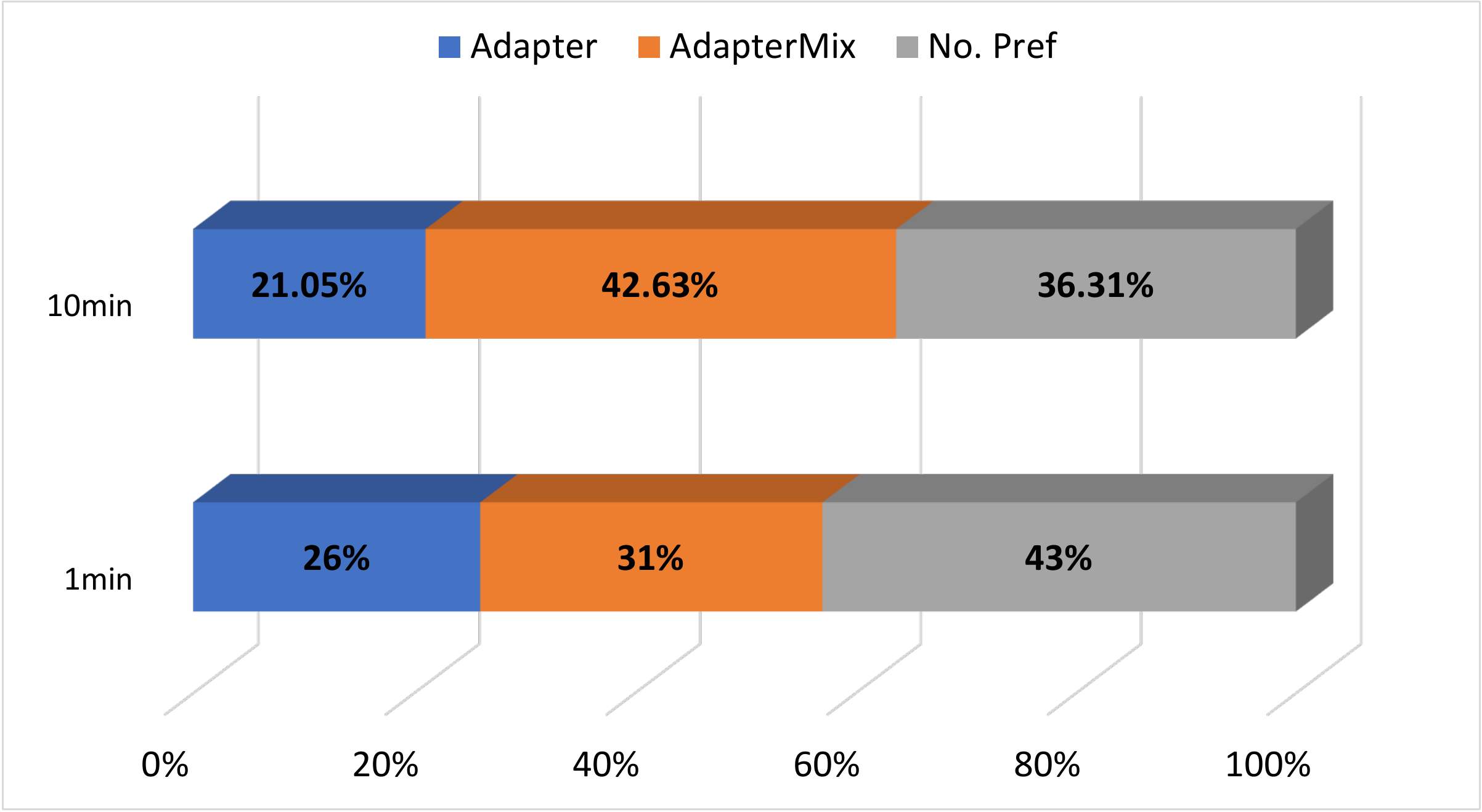}
  \caption{XAB speaker similarity test results for $1$ min and $10$ min of train data.}
  \label{fig:XAB}
\end{figure}

We conducted subjective evaluations of synthesized samples using crowdsourced Mean Opinion Score (MOS), which provides a quantitative measure of the fidelity of the synthesized audio relative to the original audio\footnote{Audio samples are available at \url{https://adaptermix.github.io}}. In this evaluation, listeners were asked which sample sounded more natural. To determine speaker individuality, we conducted an XAB test \cite{mizuno1995voice} to assess speaker similarity. The target speaker's reference speech is presented as X, and the speech synthesized by \model{} and \texttt{Adapter} are presented to listeners in a random order as A and B. Listeners were asked to choose which of A or B sounded more similar to X. A total of 20 listeners with backgrounds in NLP and speech participated in the experiment and were presented with 60 (+20 reference samples) synthesized speech samples. The evaluation samples consisted of speech samples synthesized using models trained on 1 min, 10 min, and 15 min durations. Table \ref{tab:performance} reports the MOS scores for different models, while the preference test between \model{} and \texttt{Adapter} is presented in Figure \ref{fig:XAB}.

Table \ref{tab:performance} illustrates that \model{} outperforms \texttt{Adapter}, particularly in low-resource scenarios ($1$ min and $10$ min). \model{} performs comparably or better than full fine-tuning in all three training scenarios while being significantly more parameter-efficient (approximately 10 times lower than full fine-tuning). Also, the XAB test results (Figure \ref{fig:XAB}) show that samples generated by \model{} are preferred over \model{} by $43.63\%$ and $31\%$ for training durations of $10$ min and $1$ min, respectively. Furthermore, \model{} achieved competitive or better performance in both naturalness and speaker similarity compared to the full fine-tuning under low-resource conditions. The results presented in Figure \ref{fig:rank} show the subjective ranking of speech samples synthesized by each model. Interestingly, the \model{} model achieved an average rank of $2.18$, outperforming full fine-tuning which only received a rank of $2.6$. This suggests that speech samples generated using \model{} are preferred over those generated using full fine-tuning. Notably, the \model{} only optimized $11.62\%$ of the backbone model parameters to achieve these results, showing its efficiency against full fine-tuning.

\vspace{-0.6cm}
\section{Conclusions and Future Works}
\label{sec:conclusion}
The present paper proposes a novel approach, called \model{}, for efficient TTS speaker adaptation under low-resource settings, which leverages a mixture of adapters. \model{} achieves competitive or better performance than both fine-tuning and single adapter baselines in terms of naturalness and speaker similarity. Remarkably, \model{} only optimizes $11.62\%$ of the total backbone model parameters and outperforms full model fine-tuning for low-resource speaker adaptation. 
Future extensions of this work could significantly improve the quality of synthesized speech and reduce the mismatch between TTS synthesized speech and target speaker speech. The proposed adapters and mixture of adapters provide a parameter-efficient method for adapting TTS. In future work, we could explore more challenging scenarios of extremely low-resource settings, where adaptation data is limited to less than one minute. It would also be interesting to investigate other parameter-efficient adapters, such as tiny adapters, instead of residual adapters, and explore different stochastic routing algorithms.

\section{Acknowledgement}
This project is supported by the AcRF MoE Tier-2 grant (Project no. T2MOE2008, and Grantor reference no. MOE-T2EP20220-0017) titled: “CSK-NLP: Leveraging Commonsense Knowledge for NLP”, and the SRG grant id: T1SRIS19149 titled “An Affective Multimodal Dialogue System”.




\bibliographystyle{IEEEtran}
\bibliography{main}

\end{document}